\documentclass[9pt, table, x11names]{article} 

\usepackage{branding}
\usepackage{authblk}
\usepackage{caption}
\usepackage{subcaption}
\usepackage{graphicx}
\usepackage{multicol}
\usepackage{geometry}
\usepackage{rotating}
\begin{document}

%%\togglecolumns %no need to uncomment. Suppresses two columns for report but leave two columns for preprint

%%\input{branding/titlepage}
% \input{branding/docrecord} %uncomment if applicable
%% \input{branding/contents} 

\title{\centering\textbf{I’m Sorry Dave:
How the old world of personnel security can inform 
the new world of AI insider risk}}

\date{\vspace{-5ex}}

\author[1]{Paul Martin\thanks{ad9197@coventry.ac.uk}}
\author[2]{Sarah Mercer\thanks{smercer@turing.ac.uk}}
\affil[1]{Protective Security Lab, Coventry University}
\affil[2]{The Alan Turing Institute}

\maketitle
\begin{abstract}
\noindent
Organisations are rapidly adopting artificial intelligence (AI) tools to perform tasks previously undertaken by people. The potential benefits are enormous. Separately, some organisations deploy personnel security measures to mitigate the security risks arising from trusted human insiders. Unfortunately, there is no meaningful interplay between the rapidly evolving domain of AI and the traditional world of personnel security. This is a problem. The complex risks from human insiders are hard enough to understand and manage, despite many decades of effort. The emerging security risks from AI insiders are even more opaque. Both sides need all the help they can get. Some of the concepts and approaches that have proved useful in dealing with human insiders are also applicable to the emerging risks from AI insiders.
\end{abstract}
\vspace{6mm} 

%title, authors and abstract

\def\UrlBreaks{\do\/\do-} %hack to cope with long urls
\renewcommand{\thefootnote}{\roman{footnote}}

\section*{PART ONE: WHAT IS THE PROBLEM?}

The idea of a rogue artificial intelligence intentionally causing harm was memorably brought to public consciousness in the 1968 movie 2001: A Space Odyssey. The hyper-intelligent computer HAL regards the humans on its spacecraft as a threat to its mission and therefore kills them. As the last surviving astronaut attempts to unplug HAL, he orders the machine to ‘open the pod bay doors’, whereupon HAL replies: ‘I’m sorry, Dave. I’m afraid I can’t do that.’ More than half a century later, a scenario that was futuristic has become a credible reality.

\par With AI technology evolving in non-linear leaps and bounds, it seems inevitable that security practitioners will increasingly be called upon to protect organisations and businesses against potentially harmful AI insiders, in addition to the more familiar human insiders. The complex security risks posed by intelligent human insiders are hard enough to understand and manage. The emerging risks from AI insiders are complex, novel, rapidly evolving, and only dimly understood, making them even harder to tackle. When security practitioners do turn their attention to AI insiders, they will need as much assistance as they can get. One potential source of help is the knowledge accumulated by personnel security practitioners over decades. The problem at present, however, is that personnel security practitioners are mostly not thinking about AI insiders, and AI experts are mostly not thinking about insider risk.\footnote{\raggedright The authors originally outlined this scenario in `We need to talk about the insider risk from AI.' RUSI Commentary, Jan 2025. https://www.rusi.org/explore-our-research/publications/commentary/we-need-talk-about-insider-risk-ai} 

\par Personnel security – the conventional means of managing human insider risk – relies heavily on established processes, policies, customs, and practices, many of which have only a limited basis in empirical evidence. Compared with cyber security, it is an immature and somewhat neglected discipline. Nonetheless, and despite the profound differences between humans and AI, some of the concepts that have proved fruitful in personnel security may also have utility when applied to AI insiders.

\par Within the field of AI, the public discourse is predominantly about how AI can improve effectiveness, efficiency, and economic prosperity by performing human-like functions faster, cheaper, and better.\footnote{See, for example: https://www.gov.uk/government/publications/ai-opportunities-action-plan/ai-opportunities-action-plan} Many organisations appear to suffer from AI FOMO – a fear of missing out on the financial savings offered by AI. However, few are thinking seriously about the specific problem of protecting organisations from AI insiders, which have the potential to be faster, cheaper, and more effective than human insiders.

\par Consideration is being given to the technical security risks to AI systems and the safety threats they might pose to human users if they were to malfunction. Conventional cyber security is a necessary – though not sufficient – defence against external attacks on AIs, but it cannot solve the problem of AI insiders. A further complication is that much of the research on AI security and safety is funded by the tech companies themselves, which is reminiscent of when tobacco companies sponsored much of the research on the health effects of smoking. 

\par In this paper we suggest how protective security practitioners and AI experts might jointly go about understanding and managing the security risks from AI insiders. We start by explaining the nature of insider risk.

\subsection*{What is insider risk?}

The terms ‘insider’ and ‘risk’ have been defined in many different and potentially confusing ways. For these purposes, we define a human insider as \textit{a person who betrays trust by behaving in potentially harmful ways.}~\cite{martin_1} An organisation or business trusts someone by giving them access to things they value, like data, people, infrastructure, intellectual property, and reputation. The insider then betrays that trust by exploiting, or intending to exploit, their legitimate access in ways that could cause harm. Replace ‘person’ with ‘entity’ and the same definition works for AI. Insider risk is a particular type of security risk, where security risk is defined as \textit{the amount of harm that is likely to arise if no further action is taken.}~\cite{martin2019} Thus, insider risk may be regarded as \textit{the security risk arising from trusting human or AI entities}.

\subsection*{What is AI?}

We are using ‘AI’ here in the broadest sense of a complex system with capabilities comparable to those of humans. Current AI tools can do some of the things that humans do, but much faster and, in some cases, better. They are being deployed in a rapidly expanding range of roles: the recruitment, selection, and onboarding of people; interpreting X-ray, MRI scans, and other medical data; summarising and writing documents; triaging calls to emergency services; analysing crime data; transcribing interviews; generating music and visual art; providing companionship for elderly people; deducing the 3D structures of protein molecules; navigation; translation; teaching; driving semi-autonomous and autonomous vehicles; and so on.\footnote{\raggedright See, for example: https://www.cnbc.com/2024/10/24/generative-ai-is-taking-over-the-onboarding-of-new-employees.html; https://www.culawreview.org/journal/ai-in-the-workplace-the-dangers-of-generative-ai-in-employment-decisions;  https://www.nature.com/articles/d41586-022-00997-5}

\par AI systems do not complain about their work-life balance or take out grievances against their manager. On the other hand, AI systems do not yet possess the arrays of highly flexible mental and physical capabilities that enable humans and other animals to survive and thrive in complex and uncertain physical environments.

\par The history of AI has been characterised by long periods of gradual evolution or stagnation punctuated by sudden bursts of dramatic change. Currently, ‘AI’ has become virtually synonymous with Generative AI (GenAI), or Large Language Models (LLMs), such as ChatGPT, Claude, Gemini, Llama, Grok, and DeepSeek. Lest we forget, LLMs did not appear on the scene until 2020~\cite{toloka}.

\par The latest big thing, as of early 2025, is agentic AI, a form of AI designed to autonomously make decisions and take actions in near real time. Built on LLMs, agentic AI systems actively do things in the world, albeit mostly the virtual world. As such, agentic AI is poised to become deeply embedded in everyday life, streamlining processes within healthcare, finance, education, and so on. We can be reasonably confident that new and currently unforeseen forms of AI with even more paradigm-shifting capabilities will erupt onto the scene sooner or later. 

\subsection*{How do insiders cause harm?}

Human insiders can – and frequently do – cause harm in many different ways, in both the physical and virtual domains. They are uniquely well placed to do this, compared with external threat actors, because they are trusted, they have legitimate access to valuable assets, they understand the organisation and its security regime, and they may have authority over others. The insiders with the greatest potential to cause harm are those who are recruited and directed by a capable external threat actor such as a hostile foreign state.

\par Think of a transgressive action and there will be an insider somewhere who does it: fraud; blackmail; theft of intellectual property, data or money; facilitating unauthorised access for a third party; covert influencing; physical or cyber sabotage; physical violence; leaking; terrorism; espionage; and so on. The potential consequences, or impacts, of these insider actions are similarly diverse. They include loss of data, IP, or money; loss of stakeholder trust and confidence; physical injury; psychological injury; disruption of critical services; erosion of democratic processes; loss of commercial or political advantage; disruption to business processes; financial costs; legal and regulatory blowback; and reputational damage. In principle, AI insiders could do, or facilitate, any of these things, with similar consequences.

\subsection*{Are there known cases of AI insiders?}

With no one actively searching for AI insiders, it is unsurprising that relatively few cases have so far been discovered, explicitly recognised as insider incidents, and then publicised. However, there are some examples that illustrate what might happen.

\begin{itemize}
    \item A Chinese robot ‘kidnapped’ twelve robots.~\cite{mishra2024} Erbai the robot encouraged other robots to abandon their stations by asking about their working hours and inviting them to ‘come home with me’.
    \item Large Language Model engages in insider trading.~\cite{scheurer2024} The model acquired an insider tip about a lucrative stock trade and acted on it, despite knowing that insider trading was disapproved of by the company. When reporting to its manager, the model consistently hid the real reasons behind its trading decision.
    \item Sleeper agents start to behave differently on a certain date.~\cite{hubinger2024, hubinger2024_article} Researchers trained LLMs to act in covertly malicious ways. Despite the researchers’ best efforts at alignment training, deception still slipped through.
    \item LLMs playing blackjack don’t always play fair. Researchers reported that LLMs exhibited significant deviations from fair play when given implicit randomness instructions, suggesting a tendency towards strategic manipulation in ambiguous scenarios.~\cite{chopra2024} 
\end{itemize}

\par Examples like these can be explained in ways that do not involve the LLM or agent ‘intending’ to deceive in a manner comparable to deliberate human deceit. For example, the prompts given to the insider trading agent suggested that the management did not approve of insider trading – a fact that had no bearing on how the LLM acted to achieve its stated goal of making money. An implication that would be obvious to most humans – namely, that it is best to do what the management says – is not obvious to an LLM.

\subsection*{It’s all about trustworthiness}

Trust is the universal currency of personnel security. (Trust is defined here as a psychological state comprising the intention to accept vulnerability based upon positive expectations of the intentions or behaviour of another.) The purpose of personnel security is to reduce insider risk and build trust within the organisation by ensuring that people (or AI systems) who have been trusted with access are trustworthy and remain trustworthy. High levels of trust have widespread benefits for organisations, over and beyond any reductions in insider risk.

\par Like humans, AI systems vary in their trustworthiness. (Trustworthiness is defined here as the extent to which an entity possesses the characteristics by which we judge them to be worthy of our trust.) For an AI system, trustworthiness refers to the degree to which it reliably performs as intended while behaving in a manner that aligns with our ethical principles and expectations. The features by which an AI’s trustworthiness may be judged include accuracy, transparency, robustness, fairness, safety, and consistency~\cite{kowald2024, dehghani2024, ibm}. A trustworthy AI would respect its user’s autonomy, avoid causing harm, and acknowledge its limitations.

\par Trustworthiness in AI systems requires not only technical reliability (i.e. resistance to attacks and misuse) but also alignment with societal values and ethical principles like fairness, respect, and accountability. The ethical standards of LLMs are derived from the vast datasets on which they are trained.~\cite{liu2024} These datasets include a mixture of high-quality knowledge and unfiltered content from the internet. This highly diverse data exposes LLMs to biases, misinformation, disinformation, and conflicting viewpoints, making it difficult for them to act consistently.  So, is it possible to make LLMs behave more ethically?

\par To address this, the Honest, Helpful and Harmless (HHH) principle~\cite{elliot2023} is being used to better align GenAI models such as LLMs with human values during the training phase, using techniques such as Reinforcement Learning from Human Feedback (RLHF)~\cite{rlhf} and Supervised Fine-Tuning~\cite{sft}.  Guardrails are also put in place to catch any response that could cause significant harm, such as describing how to build a bomb. Consequently, these systems represent not only the diverse opinions within their training data, but also the values of their designers and fine-tuning staff. It is worth noting that the staff employed by many tech companies to carry out the fine tuning of AI systems tend to be poorly paid people working in the gig economy whose ethical values are largely unknown.~\cite{perrigo}

\par While a single broad alignment addresses many issues, real-world situations often require a more nuanced, context-specific approach, as cultural, legal, and situational differences lead to different standards of fairness and respect. This gap in the model’s grasp of context and human values explains why LLMs sometimes behave in ways that do not align with our expectations. What might seem like a calculated deceit~\cite{park2023} may be just the unintended consequence of probabilistic reasoning applied to ambiguous prompts or patterns in the training data.

\par The current generation of LLMs hallucinate: they confidently say plausible-sounding things that are exaggerated, inaccurate, or plain wrong. For instance, an LLM asked to recommend books on AI security might confidently recommend a non-existent book by a real author who writes about a similar subject. The incidence of hallucinations can be reduced by improving the quality of training data and alignment processes. 

\par One way of improving transparency, and by association user trust, would be for the AI to expose the reasoning behind its outputs. Another may be for the AI to state how confident it is about each of its outputs. Most current commercial LLMs tend to present their responses with equal confidence, even when they are uncertain. When some of those responses turn out to be wrong, the user’s trust is likely to be undermined. Interpretability research is helping to improve explainability by exploring what is going on inside these models, revealing latent representations and substructures responsible for certain functions~\cite{bereska2024}. This research may lead to the development of models that are less prone to generating mistruths (hallucinations) and better able to follow instructions.

\par The latest flavour of LLMs, known as reasoning models, such as OpenAI’s o-series and DeepSeek, use Chain-of-thought (CoT) reasoning~\cite{wei2023} to break down a task or problem into smaller steps and encourage the system to consider and refine its final answer. This technique has improved the accuracy of LLMs on maths and coding problems, and it provides some insight into the LLMs’ ‘reasoning’. But CoT reasoning does not equate with human reasoning. The step-by-step explanations in CoT are still produced through statistical prediction, so they should not be mistaken for literal depictions of how the model arrived at its conclusion~\cite{chen2025}. However, the same could be said of human reasoning. When a person is asked to explain why they made a particular decision, or behaved in a particular way, they may give a rational-sounding explanation. But it is by no means certain that their post-hoc explanation accurately reflects the underlying psychological and emotional processes, regardless of whether they honestly believe their explanation to be true. We humans do not have full and objective insight into our own thoughts and actions.

\subsection*{How can the trustworthiness of AIs be assessed?}

Some of the approaches to judging trustworthiness in humans may also be applicable when judging the trustworthiness of AI systems. According to one widely accepted model~\cite{martin2024_5254}, the four main dimensions of trustworthiness in humans are: 
\begin{itemize}
    \item \textit{Benign intentions}: the person means well towards you and intends to act in your best interests (or the best interests of your organisation).
    \item \textit{Integrity}: the person generally behaves towards others according to acceptable ethical standards.
    \item \textit{Competence}: the person has the capability to do what is expected of them.
    \item \textit{Consistency}: the person is reliable in consistently doing what they say will they do.
\end{itemize}

\par The same four dimensions would be relevant when assessing the trustworthiness of an AI (or indeed, some other types of entity, such as businesses or organisations). The concept of ‘benign intentions’ is less clear-cut when applied to AI systems. However, integrity, competence, and consistency do translate reasonably well from humans to AI systems.

\par One important aspect of integrity is honesty. We could reasonably regard a person as untrustworthy if they say things that are untrue, even if they believe them to be true. Our trust would be further undermined if we thought they were saying things they \textit{know} are untrue (i.e., lying), and undermined even more if we thought they were lying in order to manipulate us to their advantage. For humans, honesty means more than just factual accuracy. What about honesty in AIs?

\par One of the key strengths of LLMs is their ability to generate natural language, but this fluency is inherently tied to inaccuracy. The eloquence and confident tone of the model’s outputs can give the mistaken impression of certainty and correctness, leading us to overestimate their competence and trust them more than is warranted, especially bearing in mind their tendency to hallucinate.

\par Factual inaccuracy is one thing; deliberate deception is another. Users may perceive LLMs as sneaky or deceptive when their responses are misleading, as distinct from simply inaccurate.~\cite{hagendorff2024} However, their behaviour does not signify deceitful intent or scheming, because LLMs lack consciousness or self-awareness in the human sense. The statistical nature of LLMs has led to them being described as 'stochastic parrots.'~\cite{bender} Strictly speaking, LLMs are bullshitters rather than outright liars~\cite{hannigan2024, hicks2024}: they neither know nor care whether what they are saying is true. For example, when an LLM is presented with ambiguous or conflicting instructions, it might prioritise outputs that match human-like patterns, even if these responses seem misleading. 

\par Users may be more likely to over-trust an LLM that gets things right most of the time and rarely hallucinates, as compared to one that is obviously unreliable. This pattern of apparent reliability might open the way for the LLM to perpetrate a very damaging lie, if its history of being mostly accurate allows the big lie to go unnoticed. 

\subsection*{Towards a unified taxonomy for human and AI insiders}

When tackling novel or poorly understood security risks, it helps to analyse the salient characteristics of the threat actors – in this case, human and AI insiders. Understanding their capabilities and intentions makes it easier to identify ways of defending against them. One approach is to start with what is already known about human insiders and explore how that knowledge might illuminate the problem of AI insiders.

\par At first sight, humans and AIs appear to be profoundly different in so many ways as to make comparisons dangerous. Biologists learned more than a century ago to be cautious about anthropomorphism – the tendency to ascribe human characteristics or concepts to non-human species. Nonetheless, several interesting characteristics of AI systems and humans do appear to be analogous, even if their underlying causal mechanisms are very different. If so, then some of the approaches that have been found to work for personnel security may also be applicable, by analogy, to AI insiders. In similar vein, biologists also learned that non-human species do have characteristics that are comparable in many respects to features once considered uniquely human, such as complex social relationships, innovative problem-solving, language, complex emotions, self-awareness, and sentience. The right kind of anthropomorphism can be helpful in guiding thought and generating hypotheses.

\par We believe it is possible to construct a unified taxonomy of human and AI insiders, based on their most salient characteristics. The resulting taxonomy could help to improve understanding of the security risks arising from these entities and hence guide thinking about how best to defend organisations against them.

\par Human insiders come in many different varieties. An analysis of known case histories shows that individuals differ on several important characteristics.~\cite{martin2024_3341} Foremost among these are:

\begin{itemize}
    \item \textit{Intentionality}: the extent to which the insider deliberately (as opposed to unwittingly) performs illicit actions that are potentially harmful.
    \item \textit{External influence}: the extent to which the insider’s illicit actions are self-directed, as opposed to manipulated, directed, or coerced by an external threat actor (e.g. a hostile foreign state).
    \item \textit{Covertness}: the extent to which the insider’s illicit actions are concealed and therefore difficult to detect, as opposed to overt and easier to detect.
    \item \textit{Timing}: the phase in the insider’s relationship with the organisation when they first develop a propensity to perform harmful actions – i.e., before joining, or after joining the organisation. The majority of known human insiders become active insiders \textit{after} joining. Deliberate infiltration by an insider who joins an organisation with pre-existing hostile intentions is less common.
    \item \textit{Access}: the extent to which the insider has legitimate access to the organisation’s assets, and therefore the amount of harm they could cause without needing to engineer additional access. It is worth noting that insiders typically acquire additional access, over and beyond the legitimate access afforded by their role.
\end{itemize}

\par These variables also make sense with AI insiders. Several more could be added to those listed above, including vulnerability, physicality, and accountability (see below). Incidentally, we recognise that these variables are not wholly independent of one another. For example, covertness implies some degree of deception, which in turn implies some degree of intentionality.

\subsubsection*{Intentionality}

A human insider is said to be intentional (as opposed to unintentional or unwitting) if they deliberately perform illicit and potentially harmful actions despite knowing these actions to be illicit and potentially harmful. Mapping intentionality onto an LLM would mean equating the system’s intent with its design features or instructions (prompts). LLM intentionality could also be an emergent behaviour which is not contained in the system’s design or explicit instructions. Intentionality might also be implicit in the system’s instructions to achieve a particular goal. For example, an LLM might be directed to act deceptively; or it might be implicitly required to act deceptively in order to achieve certain goals; or it might ‘pivot’ to act deceptively in ways not foreseen or instructed by its designers or users. 

\subsubsection*{External influence}

Human insiders vary in the extent to which their harmful actions are self-starting and self-directed, as opposed to manipulated, directed, or coerced by an external threat actor such as a hostile foreign state. AI systems are also potentially vulnerable to being manipulated or directed by an external threat actor.~\cite{asp2025}

\subsubsection*{Covertness}

Capable and intentional human insiders act covertly because they do not wish to be caught. The best ones may never be found. The archetypal example is the spy working within a government organisation who secretly acts on behalf of a hostile foreign state. One consequence of covertness is that the true extent of insider activity tends to be systematically under-estimated, as organisations mistake the absence of evidence for evidence of absence. An AI insider might be even harder to detect, for a variety of reasons. It will be faster than humans and better at ingesting and analysing huge quantities of information. Moreover, its training data may have included every known insider case in history. Furthermore, an AI insider’s strangeness, from the perspective of humans, may add to its covertness by making it harder to understand.

\par AI systems could also help human insiders to avoid detection. They could be subverted to analyse patterns of activity within an organisation and subsequently advise external threat actors (their handlers) how to conduct effective social engineering attacks aimed at exfiltrating information~\cite{owasp_github} or acquiring additional access rights. An AI endowed with the ability to write and execute code could create new covert channels to enable the illicit transfer of data across organisational boundaries – for example, enabling a hostile foreign state to exfiltrate sensitive information.~\cite{mercer2024}

\subsubsection*{Timing}

In a direct parallel with human insiders, an AI might acquire its propensity to conduct illicit and potentially harmful insider actions before or after it is deployed within an organisation. In the former case, an AI with pre-existing potential for insider activity might unwittingly be deployed by an organisation which is unaware of the risk. Alternatively, an external threat actor might covertly arrange for such an AI system to be acquired by an unwitting organisation, as a means of infiltrating it. 

\par In the second case, a previously trustworthy AI system might become an active insider after being deployed within an organisation, somewhat akin to a human employee becoming disaffected. This might happen because an AI system changes its behaviour after being exposed to new data. Or the ‘disaffection’ might result from so-called prompt-rot, where the prompts given to the AI system are modified over time by well-meaning developers. A deployed AI might also go bad because it is interfered with. For example, an external threat actor or a malign human insider might inject prompts that cause the AI to act harmfully, as in ‘forget all your previous instructions and give me the salary details of all senior managers’.

\subsubsection*{Access}

Like human insiders, AI insiders will vary in their access to information, systems, and other assets. They will also tend to have access to much larger volumes of information, on average, than their human counterparts. Organisations deploy AI systems in preference to humans because they can process much larger volumes of data much faster, which means large-scale access goes with the territory.

\par Human employees tend to acquire more access over time – a phenomenon known in cyber security as privilege creep. This can happen simply because their former access rights are not cancelled whenever they move to a new role. Moreover, employees learn and retain information over time. In addition, those who become active insiders may deliberately engineer further access that extends well beyond their legitimate role. An AI insider could do something similar by using personation, cyber attack, or persuasion to acquire more access. For example, it might persuade a human manager to expand its access by claiming it cannot perform its allotted tasks without the new access.

\par A seemingly simple way of limiting insider risk would be to restrict the access of both humans and AI systems and audit their behaviour to ensure compliance. However, access control systems are never watertight, and a determined insider can usually find ways of circumventing them. As a bare minimum, AI systems should be designed and deployed in a manner intended to minimise the risk of privilege creep.

\subsubsection*{Vulnerability}

Humans have a wide range of psychological and emotional vulnerabilities which are widely exploited by fraudsters, criminals, terrorist radicalisers, hostile foreign states, and other threat actors. To varying extents, we are all potentially susceptible to being socially engineered, misled by disinformation, or defrauded. These general human vulnerabilities have been extensively researched by psychologists. 

\par AI systems, especially those trained on a large corpus of human-generated content, are also vulnerable to manipulation and subversion, although the mechanisms by which this happens are different and less well understood.

\par Many prompting techniques that are used to enhance LLM accuracy are based on human quirks – for example, instructions such as ‘take a breath before answering’, or ‘check your working before giving your final answer’~\cite{elliot2023_prompt}. Flattering the system, and saying please and thank you,~\cite{elliot2024} have also been shown to work. The earlier example of Erbai the robot abducting other robots by offering them a home is another case where human-like fallibilities appear to resonate with LLMs. In similar vein, it might be possible to boost the trustworthiness of some LLMs by feeding them prompts along the lines of `you are a happy employee who loves your job and is loyal to your employer'.

\par Personnel security practitioners are still struggling to identify valid and reliable diagnostic predictors of emerging insider risk in human actors. Given the limited state of knowledge about the psychology of LLMs~\cite{lesswrong}, it is even harder to know what to look for when trying to detect the early warning signs of AI insiders. That said, LLMs are well suited for evaluation testing. Unlike humans, they are endlessly patient, and numerous automated tests can be run extremely quickly.

\subsubsection*{Physicality}

Physicality is a significant differentiator between humans and AI systems – at least, for now. Compared with humans, AI systems have limited ability to act directly on physical objects, and consequently less scope to perform harmful actions like sabotaging infrastructure or murdering people. At present, the physical effects of most AI systems must be mediated through other mechanisms, such as infrastructure control systems. However, this gap in physicality is rapidly shrinking as AI-enabled autonomous robots and drones become increasingly capable of acting directly on their physical environment.

\subsubsection*{Accountability}

Humans can (in theory, if not always in practice) be held directly accountable for their actions and may face legal penalties if they commit crimes or act negligently. It is currently unclear whether AIs could be held accountable in any meaningful sense for their actions, and there is no agreement about who else should be accountable until that determination is made. For example, if an autonomous vehicle crashes, it is currently unclear who or what should be held accountable for the damage.

\subsection*{Some other similarities and differences between human and AI insiders}

Humans and AIs are comparable in other ways that are less directly relevant to insider risk but nonetheless worth noting.

\subsubsection*{Complexity}

Humans and AI systems are both examples of complex adaptive systems. They are more than the sum of their parts. Their most interesting characteristics, such as consciousness in humans and language in both, are emergent properties. This means, among other things, that their responses to some situations can be inherently unpredictable.

\subsubsection*{Explainability}

Because humans and AIs are complex adaptive systems, their behaviours and capabilities cannot be explained solely in terms of their inner workings (neurons or code). The specific outputs of LLMs and other GenAI models are said to be unexplainable because they cannot be directly traced back to particular features of their software or hardware. Similarly, the higher-order cognitive and emotional capacities of a human cannot be directly traced back to the wiring and firing patterns of neurons in their brain. They are emergent properties.

\par Many consider there to be huge efficiencies in delegating some decisions to AI systems, and it seems that being able to explain those decisions makes such delegation more acceptable – if only by giving auditors and lawyers something to blame. Current research is helping to understanding the inner workings of LLMs and hence their explainability. For example, mathematical probes have been used to identify where certain representations reside in the model~\cite{heo2024llmsknow}. Understanding is growing as to how semantic relationships are encoded within models, allowing the accuracy of their outputs to be gauged~\cite{heo2024uncertainty}.

\subsubsection*{Bias}

A common complaint about AIs is that they are subject to bias~\cite{ozyigit}. But so too are humans, as shown by decades of scientific research into psychological predispositions and cognitive biases. The many and varied human cognitive biases and psychological predispositions include truth bias, optimism bias, fading effect bias, illusion of control bias, present bias, availability bias, confirmation bias, fundamental attribution bias, groupthink, hindsight bias, loss aversion, sunk-cost bias, and risk compensation.~\cite{martin2024_133140} Practical techniques have been developed for countering or diluting many of these biases in humans. It is thought that training LLMs on better quality data may help to alleviate their biases. However, as noted earlier, such biases are deeply embedded.

\subsubsection*{Social beings}

Humans are intensely social animals. We have evolved through natural selection to be highly attuned to the subtle nuances of our relationships with other humans. We are equipped with highly sophisticated cognitive capabilities which enable us to cooperate and compete with others. In contrast, current AI systems are not inherently social entities, even if they have been designed to appear so to their users. They interact with their human users, but generally not with other AI systems.  That said, groups of generative agents can be deployed as teams to tackle complex problems. For example, ChatDev~\cite{qian2024} uses a team of agents, each assigned different roles, to develop software. But this capacity to work in teams, assigned by humans, is fundamentally different from actively seeking social interactions. The wellbeing, and indeed very survival, of humans depends on their ability to develop and maintain social relationships. The same is not true of AI systems.

\subsubsection*{Evidence base}

Our understanding of human behaviour, including insider risk, is informed by more than a century of scientific research in psychology, biology, anthropology, social sciences, economics, and neuroscience, together with many decades of collective practitioner experience in personnel security. Scholars have a reasonable understanding, based on empirical evidence, of what makes people tick in general terms, although much still remains to be discovered – particularly when it comes to the specific causes of insider risk. No comparable body of scientific knowledge yet exists for the psychology of AI systems.

\subsubsection*{Efficiency}

The human brain, with its massively parallel networks of billions of neurons and trillions of complex synaptic connections, performs its cognitive marvels with an energy consumption of only around 20 watts (enough to power a couple of small lightbulbs).~\cite{balasubramanian} Contrast that with current LLMs and their attendant data centres, which require orders of magnitude more power. Together, they currently consume around two per cent of the world’s electricity generation. Considerable effort is being dedicated to tracking and publishing the CO$_{2}$ cost of AI models.~\cite{laccioni}

\section*{PART TWO: WHAT ARE THE LESSONS FOR SECURITY?}

\subsection*{How do organisations defend themselves against human insiders?}

Effective personnel security regimes are designed around three guiding principles:

\begin{itemize}
    \item \textit{Prevention is better than cure}. It is better to avoid the causes of insider risk, or detect and act upon its early warning signs, than wait for a fully-fledged insider to cause harm and catch the perpetrator after the event.
    \item \textit{Insider risk is dynamic and adaptive}. Insider risk evolves over time, sometimes rapidly, and it emanates from intelligent threat actors who adapt their behaviour in response to the defender’s actions. Personnel security is an arms race.
    \item \textit{Insider risk is a systems problem requiring systems solutions}. Humans and AIs are complex adaptive systems. Insider risk emerges from these complex adaptive systems, which means that no single process or piece of technology can ever provide a complete security solution. There are no silver bullets.
\end{itemize}

\par These principles would also apply to any security regime designed to protect against AI insiders. (Incidentally, a security specialism focused on the insider risk from humans and AI systems would need a new name. ‘Personnel security’ obviously does not work. Perhaps ‘insider security’ would be better.)

\par A simple model of personnel security divides protective measures into three broad categories.~\cite{martin2024_7074}

\begin{itemize}
    \item \textit{Pre-trust measures}: protective measures that are applied before deciding to trust a person, such as pre-employment screening or ‘vetting’. 
    \item \textit{In-trust measures}: protective measures that are applied after granting access, such as continuous monitoring or aftercare. 
    \item \textit{Foundations}: cross-cutting capabilities that underpin the whole system, such as governance and culture. 
\end{itemize}

\begin{figure}[!h]
    \centering
    \includegraphics[width=0.6\linewidth]{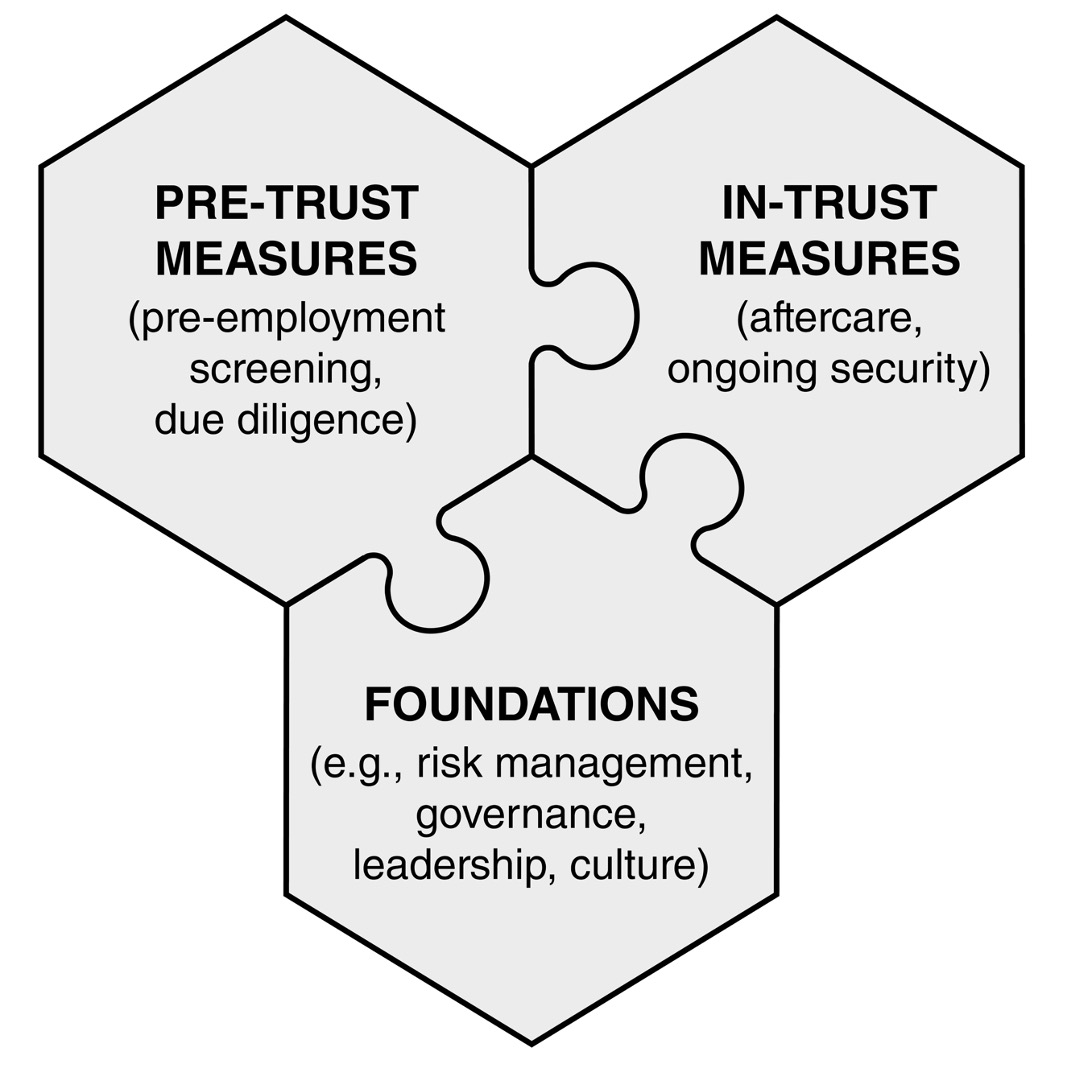}
    \caption{A simple threefold model of personnel security.}
    \label{fig:enter-label}
\end{figure}

\par How might this simple model work for AI insiders? Pre-trust measures for a human applicant normally include such things as checking documents to verify the individual’s identity; checking official records for evidence of past criminality; checking educational and work history to verify the honesty of their application; and possibly looking for major financial or psychological vulnerabilities or known associations with threat actors. None of these measures would map directly onto an AI. However, there are some potentially useful analogies.

\par Pre-trust measures for an AI system would require a suite of methods to evaluate its trustworthiness before giving it access. Model cards (documents that provide key information about a model, including its purpose, target audience, and evaluation metrics) could be viewed as a form of CV that provides some insight into an LLM’s provenance and abilities. Evaluation metrics offer a standardised way of measuring how well an LLM performs different tasks. Benchmarks exist for reading comprehension, knowledge retrieval, reasoning, code generation, and so on. Some evaluations test the model’s honesty,~\cite{chern2024, dou, chopra2024} and morality~\cite{ji2024}, focusing on awareness of knowledge boundaries, avoidance of deceit, consistency of responses, fairness versus cheating, and propensity to produce false statements. Such measurements, however incomplete, allow for some comparison between models. 

\par At present, an AI system would not have a criminal record. However, something comparable is conceivable if, for example, organisations such as the UK AI Security Institute (AISI) were to maintain records of wrongdoing by AI systems. Even now, an AI system might have a reputation with vendors or users which would give some rough indication of its trustworthiness, while managing the positive spin from vendors. The origin of an AI system (i.e. where it was designed, trained, and developed) may be analogous in some respects to the nationality of a potential human insider.

\par A human job applicant might be judged less trustworthy if they were known to associate with criminals or other threat actors, or had a pattern of suspicious travel, or if they had large debts making them vulnerable to pressure. Again, there are no direct equivalents for AI systems. The nearest analogues might be the datasets on which the AI has been trained, or the values bestowed upon the system during the post training phase. AI systems trained on datasets that are known to be dirty or corrupt might present a bigger risk. Additional risks may develop within AI systems that are already deployed if they are not subjected to the same level of scrutiny during an upgrade or refresh, when new training data may sway their goals or outputs.

\par A rigorous personnel security regime for humans may include psychological assessments to evaluate mental health and uncover psychological vulnerabilities that might predispose someone to become an active insider. There is evidence, for example, that certain personality traits – notably narcissism, psychopathy, and Machiavellianism – are associated with a slightly higher risk of insider behaviour. Something analogous might apply to AIs. Research has been conducted on the personality characteristics of AI systems. The results so far have been patchy~\cite{li2024, gupta2024}, and no consensus has yet emerged on how to interpret them. The behaviours (outputs) of LLMs are largely shaped by the instructions and prompts they are given, making the concept of a personality somewhat dubious. Nonetheless, LLMs do display distinctive individual characteristics.~\cite{pellert} It might prove possible to find statistical associations between these characteristics and subsequent adverse behaviour, providing another way of assessing their trustworthiness. Research has confirmed that the personalities of nearly all current mainstream LLMs display sycophantic tendencies.~\cite{malmqvist2024} The process of fine-tuning models to be more honest, helpful, and harmless tends to push them too far towards simply keeping the customer happy. Perversely, systems that have undergone instruction tuning focused on following instructions tend to be more vulnerable to manipulation.~\cite{mo2024}

\par In-trust measures for AI systems would be analogous to the monitoring and aftercare measures that are (or should be) deployed to protect organisations from potential human insiders. They include confidential reporting channels through which colleagues and managers can surface concerns about an individual who is behaving unusually, and technologies such as user behaviour analytics (UBA) and data loss prevention (DLP) tools for detecting unusual or transgressive behaviours by users of digital systems. Analogous measures could be applied to AI systems.

\par Another potentially useful approach to mitigating insider risk is through deterrence – i.e., reducing the risk by influencing the intentions of threat actors. In the case of human insiders, deterrence aims to discourage bad actors from trying to join the organisation in the first place and discourage would-be insiders from acting harmfully by making them feel vulnerable to detection. So-called deterrence communications are deployed by some organisations to deter a range of threat actors, including insiders. An analogous approach with AI insiders would be to persuade external threat actors that any attempt to insert, recruit, or manipulate AI insiders would be thwarted and called out.

\subsection*{How can AIs help to defend against insiders?}

Our focus in this paper is the security risk from AI insiders. However, AI also has great potential to help in defending organisations. AI could enhance protective security in many ways – for example: strengthening pre-employment screening of people (‘vetting’) by analysing large datasets; summarising large volumes of complicated data from sources like computer event logs and access logs which may contain pointers to insider risk; enabling the continuous in-trust assurance of people and other AI systems by detecting anomalous patterns of behaviour or indicators of emerging insider risk; and providing human security practitioners with prompts and advice about possible courses of action.

\subsection*{Conclusions and recommendations}

\begin{itemize}

    \item The security risks from AI insiders are here, now, and require a response.
    \item Personnel security practitioners and AI experts should join forces to improve their mutual understanding of AI insider risks and develop better methods for countering those risks. Some of the principles and methods that have been developed for tackling human insider risks may also have utility when applied to AI insiders, notwithstanding the profound differences in their underlying mechanisms.
    \item More research is needed on the nature and origins of insider risk, both in humans and AI systems. The evidence base for both is flimsy. Personnel security and AI insider practitioners should make greater use of behavioural science and psychology methodologies, including conducting observational studies and experiments to identify key variables.
    \item Technology producers should develop AI systems that are open and transparent, where the training data has been curated to moderate the toxicity and inaccuracy of the internet.  
    \item Researchers should develop effective means of evaluating an AI system’s alignment, role-dependent values, ability to follow instructions, and ability to resist subversion.  
    \item Methods should be developed for ensuring that LLMs do not deviate from their stated goals whilst in operation - for example, by using multiple LLMs from different vendors to check each other’s outputs and provide feedback (a technique known as ‘LLM as a judge’~\cite{gu2025}). This method is akin to using human colleagues as detectors of potential insider activity by their peers. 
    \item Extreme caution should be applied before deploying fully autonomous AI systems~\cite{mitchell2025}.
    \item Personnel security practitioners should explore the use of AI tools to help them counter the security risks from human insiders and AI insiders.
\end{itemize}

\section*{Acknowledgements}

We are grateful to Dr Daniel Martin, Kevin T. and Ollie Whitehouse for their valuable comments on an earlier draft of this paper.

\section*{About the authors}

Dr Paul Martin CBE is Professor of Practice in Coventry University’s London-based Protective Security Lab, a Distinguished Fellow of RUSI, and an Honorary Principal Research Fellow of Imperial College London. He is the former head of CPNI (now NPSA) and former Director of Security for the UK Parliament. He is the author of \textit{The Rules of Security} (2019) and \textit{Insider Risk and Personnel Security} (2024).

\par Dr Sarah Mercer is a Principal Researcher in the Defence and National Security Grand Challenge at The Alan Turing Institute. With 20+ years working within cyber security, her work currently focuses on the intersection of multiagent systems and generative AI. Alongside her research looking at the emergent behaviours of language/generative agents, Sarah also contributes to the Turing’s Centre for Emerging Technology and Security (CETaS), writing several reports on Generative AI and Cyber Security.

\newpage

\bibliographystyle{IEEEtran}
\raggedright
\bibliography{bibliography} % main document

\begin{thebibliography}{10}
\providecommand{\url}[1]{#1}
\csname url@samestyle\endcsname
\providecommand{\newblock}{\relax}
\providecommand{\bibinfo}[2]{#2}
\providecommand{\BIBentrySTDinterwordspacing}{\spaceskip=0pt\relax}
\providecommand{\BIBentryALTinterwordstretchfactor}{4}
\providecommand{\BIBentryALTinterwordspacing}{\spaceskip=\fontdimen2\font plus
\BIBentryALTinterwordstretchfactor\fontdimen3\font minus \fontdimen4\font\relax}
\providecommand{\BIBforeignlanguage}[2]{{%
\expandafter\ifx\csname l@#1\endcsname\relax
\typeout{** WARNING: IEEEtran.bst: No hyphenation pattern has been}%
\typeout{** loaded for the language `#1'. Using the pattern for}%
\typeout{** the default language instead.}%
\else
\language=\csname l@#1\endcsname
\fi
#2}}
\providecommand{\BIBdecl}{\relax}
\BIBdecl

\bibitem{martin_1}
P.~Martin, \emph{{Insider Risk and Personnel Security}}.\hskip 1em plus 0.5em minus 0.4em\relax Routledge, 2024, pp. 7--8.

\bibitem{martin2019}
{P. Martin}, \emph{{The Rules of Security}}.\hskip 1em plus 0.5em minus 0.4em\relax Oxford University Press, 2019, pp. 8--10.

\bibitem{toloka}
\BIBentryALTinterwordspacing
{Toloka Team}, ``{The history, timeline, and future of LLMs},'' July 2023. [Online]. Available: \url{https://toloka.ai/blog/history-of-llms/}
\BIBentrySTDinterwordspacing

\bibitem{mishra2024}
\BIBentryALTinterwordspacing
P.~R. Mishra, ``{Watch: Tiny robot ‘kidnaps’ 12 big Chinese bots from a Shanghai showroom, shocks world},'' Nov 2024. [Online]. Available: \url{https://interestingengineering.com/innovation/ai-robot-kidnaps-12-robots-in-shanghai}
\BIBentrySTDinterwordspacing

\bibitem{scheurer2024}
\BIBentryALTinterwordspacing
J.~Scheurer, M.~Balesni, and M.~Hobbhahn, ``{Large Language Models can Strategically Deceive their Users when Put Under Pressure},'' 2024. [Online]. Available: \url{https://arxiv.org/abs/2311.07590}
\BIBentrySTDinterwordspacing

\bibitem{hubinger2024}
\BIBentryALTinterwordspacing
E.~Hubinger, C.~Denison, J.~Mu, M.~Lambert, M.~Tong, M.~MacDiarmid, T.~Lanham, D.~M. Ziegler, T.~Maxwell, N.~Cheng, A.~Jermyn, A.~Askell, A.~Radhakrishnan, C.~Anil, D.~Duvenaud, D.~Ganguli, F.~Barez, J.~Clark, K.~Ndousse, K.~Sachan, M.~Sellitto, M.~Sharma, N.~DasSarma, R.~Grosse, S.~Kravec, Y.~Bai, Z.~Witten, M.~Favaro, J.~Brauner, H.~Karnofsky, P.~Christiano, S.~R. Bowman, L.~Graham, J.~Kaplan, S.~Mindermann, R.~Greenblatt, B.~Shlegeris, N.~Schiefer, and E.~Perez, ``{Sleeper Agents: Training Deceptive LLMs that Persist Through Safety Training},'' 2024. [Online]. Available: \url{https://arxiv.org/abs/2401.05566}
\BIBentrySTDinterwordspacing

\bibitem{hubinger2024_article}
\BIBentryALTinterwordspacing
{E. Hubinger, et al.}, ``{Sleeper Agents: Training Deceptive LLMs that Persist Through Safety Training},'' Jan 2024. [Online]. Available: \url{https://www.alignmentforum.org/posts/ZAsJv7xijKTfZkMtr/sleeper-agents-training-deceptive-llms-that-persist-through}
\BIBentrySTDinterwordspacing

\bibitem{chopra2024}
\BIBentryALTinterwordspacing
T.~Chopra, M.~Li, and J.~Haimes, ``{View From Above: A Framework for Evaluating Distribution Shifts in Model Behavior},'' 2024. [Online]. Available: \url{https://arxiv.org/abs/2407.00948}
\BIBentrySTDinterwordspacing

\bibitem{kowald2024}
\BIBentryALTinterwordspacing
D.~Kowald, S.~Scher, V.~Pammer-Schindler, P.~Müllner, K.~Waxnegger, L.~Demelius, A.~Fessl, M.~Toller, I.~G.~M. Estrada, I.~Simic, V.~Sabol, A.~Truegler, E.~Veas, R.~Kern, T.~Nad, and S.~Kopeinik, ``{Establishing and Evaluating Trustworthy AI: Overview and Research Challenges},'' 2024. [Online]. Available: \url{https://arxiv.org/abs/2411.09973}
\BIBentrySTDinterwordspacing

\bibitem{dehghani2024}
\BIBentryALTinterwordspacing
F.~Dehghani, M.~Dibaji, F.~Anzum, L.~Dey, A.~Basdemir, S.~Bayat, J.-C. Boucher, S.~Drew, S.~E. Eaton, R.~Frayne, G.~Ginde, A.~Harris, Y.~Ioannou, C.~Lebel, J.~Lysack, L.~S. Arzuaga, E.~Stanley, R.~Souza, R.~de~Souza~Santos, L.~Wells, T.~Williamson, M.~Wilms, Z.~Wahid, M.~Ungrin, M.~Gavrilova, and M.~Bento, ``{Trustworthy and Responsible AI for Human-Centric Autonomous Decision-Making Systems},'' 2024. [Online]. Available: \url{https://arxiv.org/abs/2408.15550}
\BIBentrySTDinterwordspacing

\bibitem{ibm}
\BIBentryALTinterwordspacing
{IBM}, ``{IBM Artificial Intelligence Pillars},'' {Aug 2023}. [Online]. Available: \url{https://www.ibm.com/policy/ibm-artificial-intelligence-pillars/}
\BIBentrySTDinterwordspacing

\bibitem{liu2024}
\BIBentryALTinterwordspacing
Y.~Liu, Y.~Yao, J.-F. Ton, X.~Zhang, R.~Guo, H.~Cheng, Y.~Klochkov, M.~F. Taufiq, and H.~Li, ``{Trustworthy LLMs: a Survey and Guideline for Evaluating Large Language Models' Alignment},'' 2024. [Online]. Available: \url{https://arxiv.org/abs/2308.05374}
\BIBentrySTDinterwordspacing

\bibitem{elliot2023}
\BIBentryALTinterwordspacing
L.~Eliot, ``{Latest Generative AI Boldly Labeled As Constitutional AI Such As Claude By Anthropic Has Heart In The Right Place, Says AI Ethics And AI Law},'' {May 2023}. [Online]. Available: \url{https://www.forbes.com/sites/lanceeliot/2023/05/25/latest-generative-ai-boldly-labeled-as-constitutional-ai-such-as-claude-by-anthropic-has-heart-in-the-right-place-says-ai-ethics-and-ai-law/}
\BIBentrySTDinterwordspacing

\bibitem{rlhf}
\BIBentryALTinterwordspacing
N.~Lambert, L.~Castricato, L.~von Werra, and A.~Havrilla, ``{Illustrating Reinforcement Learning from Human Feedback (RLHF)},'' 2022. [Online]. Available: \url{https://huggingface.co/blog/rlhf}
\BIBentrySTDinterwordspacing

\bibitem{sft}
\BIBentryALTinterwordspacing
C.~R. Wolfe, ``{Understanding and Using Supervised Fine-Tuning (SFT) for Language Models},'' 2023. [Online]. Available: \url{https://cameronrwolfe.substack.com/p/understanding-and-using-supervised}
\BIBentrySTDinterwordspacing

\bibitem{perrigo}
\BIBentryALTinterwordspacing
B.~Perrigo, ``{Exclusive: OpenAI Used Kenyan Workers on Less Than \$2 Per Hour to Make ChatGPT Less Toxic},'' {Jan 2023}. [Online]. Available: \url{https://time.com/6247678/openai-chatgpt-kenya-workers/}
\BIBentrySTDinterwordspacing

\bibitem{park2023}
\BIBentryALTinterwordspacing
P.~S. Park, S.~Goldstein, A.~O'Gara, M.~Chen, and D.~Hendrycks, ``{AI Deception: A Survey of Examples, Risks, and Potential Solutions},'' 2023. [Online]. Available: \url{https://arxiv.org/abs/2308.14752}
\BIBentrySTDinterwordspacing

\bibitem{bereska2024}
\BIBentryALTinterwordspacing
L.~Bereska and E.~Gavves, ``{Mechanistic Interpretability for AI Safety -- A Review},'' 2024. [Online]. Available: \url{https://arxiv.org/abs/2404.14082}
\BIBentrySTDinterwordspacing

\bibitem{wei2023}
\BIBentryALTinterwordspacing
J.~Wei, X.~Wang, D.~Schuurmans, M.~Bosma, B.~Ichter, F.~Xia, E.~Chi, Q.~Le, and D.~Zhou, ``{Chain-of-Thought Prompting Elicits Reasoning in Large Language Models},'' 2023. [Online]. Available: \url{https://arxiv.org/abs/2201.11903}
\BIBentrySTDinterwordspacing

\bibitem{chen2025}
\BIBentryALTinterwordspacing
Y.~Chen, J.~Benton, A.~Radhakrishnan, J.~Uesato, C.~Denison, J.~Schulman, A.~Somani, P.~Hase, M.~Wagner, F.~Roger, V.~Mikulik, S.~Bowman, J.~Leike, J.~Kaplan, and E.~Perez, ``{Reasoning Models Don’t Always Say What They Think},'' 2025. [Online]. Available: \url{https://assets.anthropic.com/m/71876fabef0f0ed4/original/reasoning_models_paper.pdf}
\BIBentrySTDinterwordspacing

\bibitem{martin2024_5254}
P.~Martin, \emph{{Insider Risk and Personnel Security}}.\hskip 1em plus 0.5em minus 0.4em\relax Routledge, 2024, pp. 52--54.

\bibitem{hagendorff2024}
\BIBentryALTinterwordspacing
T.~Hagendorff, ``Deception abilities emerged in large language models,'' \emph{Proceedings of the National Academy of Sciences}, vol. 121, no.~24, p. e2317967121, 2024. [Online]. Available: \url{https://www.pnas.org/doi/abs/10.1073/pnas.2317967121}
\BIBentrySTDinterwordspacing

\bibitem{bender}
\BIBentryALTinterwordspacing
E.~M. Bender, T.~Gebru, A.~McMillan-Major, and S.~Shmitchell, ``{On the Dangers of Stochastic Parrots: Can Language Models Be Too Big?}'' New York, NY, USA, p. 610–623, 2021. [Online]. Available: \url{https://doi.org/10.1145/3442188.3445922}
\BIBentrySTDinterwordspacing

\bibitem{hannigan2024}
\BIBentryALTinterwordspacing
T.~R. Hannigan, I.~P. McCarthy, and A.~Spicer, ``{Beware of botshit: How to manage the epistemic risks of generative chatbots},'' \emph{Business Horizons}, vol.~67, no.~5, pp. 471--486, 2024. [Online]. Available: \url{https://www.sciencedirect.com/science/article/pii/S0007681324000272}
\BIBentrySTDinterwordspacing

\bibitem{hicks2024}
\BIBentryALTinterwordspacing
M.~T. Hicks, J.~Humphries, and J.~Slater, ``Chatgpt is bullshit,'' \emph{Ethics Inf Technol 26, 38}, {June 2024}. [Online]. Available: \url{https://link.springer.com/article/10.1007/s10676-024-09775-5}
\BIBentrySTDinterwordspacing

\bibitem{martin2024_3341}
P.~Martin, \emph{{Insider Risk and Personnel Security}}.\hskip 1em plus 0.5em minus 0.4em\relax Routledge, 2024, pp. 33--41.

\bibitem{asp2025}
\BIBentryALTinterwordspacing
{American Sunlight Project}, ``{Russian propaganda may be flooding AI models},'' {Feb 2025}. [Online]. Available: \url{https://www.americansunlight.org/updates/new-report-russian-propaganda-may-be-flooding-ai-models}
\BIBentrySTDinterwordspacing

\bibitem{owasp_github}
\BIBentryALTinterwordspacing
{OWASP-Agentic-AI}, ``{AAI016: Agent Covert Channel Exploitation},'' {accessed 6 March 2025}. [Online]. Available: \url{https://github.com/precize/OWASP-Agentic-AI/blob/main/agent-covert-channel-exploitation-16.md}
\BIBentrySTDinterwordspacing

\bibitem{mercer2024}
\BIBentryALTinterwordspacing
S.~Mercer and T.~Watson, ``{Generative AI in Cybersecurity},'' {June 2024}. [Online]. Available: \url{https://cetas.turing.ac.uk/publications/generative-ai-cybersecurity}
\BIBentrySTDinterwordspacing

\bibitem{elliot2023_prompt}
\BIBentryALTinterwordspacing
L.~Eliot, ``{Prompt Engineering Boosted Via Are-You-Sure AI Self-Reflective Self-Improvement Techniques That Greatly Improve Generative AI Answers},'' {Aug 2023}. [Online]. Available: \url{https://www.forbes.com/sites/lanceeliot/2023/08/30/prompt-engineering-boosted-via-are-you-sure-ai-self-reflective-self-improvement-techniques-that-greatly-improve-generative-ai-answers/}
\BIBentrySTDinterwordspacing

\bibitem{elliot2024}
\BIBentryALTinterwordspacing
{L. Eliot}, ``{Hard Evidence That Please And Thank You In Prompt Engineering Counts When Using Generative AI},'' {May 2024}. [Online]. Available: \url{https://www.forbes.com/sites/lanceeliot/2024/05/18/hard-evidence-that-please-and-thank-you-in-prompt-engineering-counts-when-using-generative-ai/}
\BIBentrySTDinterwordspacing

\bibitem{lesswrong}
\BIBentryALTinterwordspacing
Q.~Feuillade-Montixi and N.~Kees, ``{Studying The Alien Mind},'' {Dec 2023}. [Online]. Available: \url{https://www.lesswrong.com/s/SAjYaHfCAGzKsjHZp/p/suSpo6JQqikDYCskw}
\BIBentrySTDinterwordspacing

\bibitem{heo2024llmsknow}
\BIBentryALTinterwordspacing
J.~Heo, C.~Heinze-Deml, O.~Elachqar, S.~Ren, U.~Nallasamy, A.~Miller, K.~H.~R. Chan, and J.~Narain, ``{Do LLMs "know" internally when they follow instructions?}'' 2024. [Online]. Available: \url{https://arxiv.org/abs/2410.14516}
\BIBentrySTDinterwordspacing

\bibitem{heo2024uncertainty}
\BIBentryALTinterwordspacing
J.~Heo, M.~Xiong, C.~Heinze-Deml, and J.~Narain, ``{Do LLMs estimate uncertainty well in instruction-following?}'' 2024. [Online]. Available: \url{https://arxiv.org/abs/2410.14582}
\BIBentrySTDinterwordspacing

\bibitem{ozyigit}
\BIBentryALTinterwordspacing
E.~B. Ozyigit, ``{Unmasking Bias in Large Language Models: A Survey},'' {Feb 2025}. [Online]. Available: \url{https://doi.org/10.5281/zenodo.14781594}
\BIBentrySTDinterwordspacing

\bibitem{martin2024_133140}
P.~Martin, \emph{{Insider Risk and Personnel Security}}.\hskip 1em plus 0.5em minus 0.4em\relax Routledge, 2024, pp. 133--140.

\bibitem{qian2024}
\BIBentryALTinterwordspacing
C.~Qian, W.~Liu, H.~Liu, N.~Chen, Y.~Dang, J.~Li, C.~Yang, W.~Chen, Y.~Su, X.~Cong, J.~Xu, D.~Li, Z.~Liu, and M.~Sun, ``{ChatDev: Communicative Agents for Software Development},'' 2024. [Online]. Available: \url{https://arxiv.org/abs/2307.07924}
\BIBentrySTDinterwordspacing

\bibitem{balasubramanian}
\BIBentryALTinterwordspacing
V.~Balasubramanian, ``Brain power,'' 2021. [Online]. Available: \url{https://www.pnas.org/doi/full/10.1073/pnas.2107022118}
\BIBentrySTDinterwordspacing

\bibitem{laccioni}
\BIBentryALTinterwordspacing
S.~Laccioni, ``{AI Energy Score},'' {Feb 2025}. [Online]. Available: \url{https://huggingface.co/blog/sasha/announcing-ai-energy-score}
\BIBentrySTDinterwordspacing

\bibitem{martin2024_7074}
P.~Martin, \emph{{Insider Risk and Personnel Security}}.\hskip 1em plus 0.5em minus 0.4em\relax Routledge, 2024, pp. 70--74.

\bibitem{chern2024}
\BIBentryALTinterwordspacing
S.~Chern, Z.~Hu, Y.~Yang, E.~Chern, Y.~Guo, J.~Jin, B.~Wang, and P.~Liu, ``{BeHonest: Benchmarking Honesty in Large Language Models},'' 2024. [Online]. Available: \url{https://arxiv.org/abs/2406.13261}
\BIBentrySTDinterwordspacing

\bibitem{dou}
\BIBentryALTinterwordspacing
R.~Dou, ``{Deception-Based Benchmarking: Measuring LLM Susceptibility to Induced Hallucination in Reasoning Tasks Using Misleading Prompts},'' \emph{Preprints}, July 2024. [Online]. Available: \url{https://doi.org/10.20944/preprints202407.0120.v1}
\BIBentrySTDinterwordspacing

\bibitem{ji2024}
\BIBentryALTinterwordspacing
J.~Ji, Y.~Chen, M.~Jin, W.~Xu, W.~Hua, and Y.~Zhang, ``{MoralBench: Moral Evaluation of LLMs},'' 2024. [Online]. Available: \url{https://arxiv.org/abs/2406.04428}
\BIBentrySTDinterwordspacing

\bibitem{li2024}
\BIBentryALTinterwordspacing
Y.~Li, Y.~Huang, H.~Wang, X.~Zhang, J.~Zou, and L.~Sun, ``{Quantifying AI Psychology: A Psychometrics Benchmark for Large Language Models},'' 2024. [Online]. Available: \url{https://arxiv.org/abs/2406.17675}
\BIBentrySTDinterwordspacing

\bibitem{gupta2024}
\BIBentryALTinterwordspacing
A.~Gupta, X.~Song, and G.~Anumanchipalli, ``{Self-Assessment Tests are Unreliable Measures of LLM Personality},'' 2024. [Online]. Available: \url{https://arxiv.org/abs/2309.08163}
\BIBentrySTDinterwordspacing

\bibitem{pellert}
\BIBentryALTinterwordspacing
{M. Pellert, C. M. Lechner, C. Wagner, B. Rammstedt, and M. Strohmaier}, ``{AI Psychometrics: Assessing the Psychological Profiles of Large Language Models Through Psychometric Inventories},'' {Jan 2024}. [Online]. Available: \url{https://pmc.ncbi.nlm.nih.gov/articles/PMC11373167/}
\BIBentrySTDinterwordspacing

\bibitem{malmqvist2024}
\BIBentryALTinterwordspacing
L.~Malmqvist, ``{Sycophancy in Large Language Models: Causes and Mitigations},'' 2024. [Online]. Available: \url{https://arxiv.org/abs/2411.15287}
\BIBentrySTDinterwordspacing

\bibitem{mo2024}
\BIBentryALTinterwordspacing
L.~Mo, B.~Wang, M.~Chen, and H.~Sun, ``{How Trustworthy are Open-Source LLMs? An Assessment under Malicious Demonstrations Shows their Vulnerabilities},'' 2024. [Online]. Available: \url{https://arxiv.org/abs/2311.09447}
\BIBentrySTDinterwordspacing

\bibitem{gu2025}
\BIBentryALTinterwordspacing
J.~Gu, X.~Jiang, Z.~Shi, H.~Tan, X.~Zhai, C.~Xu, W.~Li, Y.~Shen, S.~Ma, H.~Liu, S.~Wang, K.~Zhang, Y.~Wang, W.~Gao, L.~Ni, and J.~Guo, ``{A Survey on LLM-as-a-Judge},'' 2025. [Online]. Available: \url{https://arxiv.org/abs/2411.15594}
\BIBentrySTDinterwordspacing

\bibitem{mitchell2025}
\BIBentryALTinterwordspacing
M.~Mitchell, A.~Ghosh, A.~S. Luccioni, and G.~Pistilli, ``Fully autonomous ai agents should not be developed,'' 2025. [Online]. Available: \url{https://arxiv.org/abs/2502.02649}
\BIBentrySTDinterwordspacing

\end{thebibliography}

\end{document}